\documentclass[bibyear]{aa}

\usepackage{graphicx}
\usepackage{txfonts}
\usepackage{lipsum}
\usepackage{natbib} % for \citep and \citet commands
\usepackage{amssymb} % for \lesssim
\usepackage{booktabs}
\usepackage{subcaption}         % necessary for continued figures, example in section 3
\usepackage{siunitx}            % for S column type in tables
\usepackage{xcolor}             % for \textcolor command
\usepackage{colortbl}           % for \rowcolor in tables
\usepackage{bm}                       % and appendix
\usepackage{lscape}             % to rotate a single page table, example in appendix.
\usepackage{placeins}           % useful with \FloatBarrier, to keep                              % onecolumn floats from drifting to the next section
                                
\newcommand{\bh}[1]{\textbf{\boldmath #1}}

\begin{document}

\title{Constraints on the central engine of merger-driven long gamma-ray bursts}

\subtitle{}

\author{
Muskan Yadav\inst{1}\thanks{E-mail: muskan.yadav@students.uniroma2.eu}
\and Roberto Ricci\inst{1,2}
\and Paz Beniamini\inst{3}
\and Hira Waseem\inst{1}
\and Tatsuya Matsumoto\inst{4}
\and Simone Dichiara\inst{5}
\and Eleonora Troja\inst{1}
}

\institute{
$^{1}$Dipartimento di Fisica, Universit\`a di Tor Vergata, Via della Ricerca Scientifica, 1, 00133 Rome, Italy\\
$^{2}$INAF–Istituto di Radioastronomia, Via Gobetti, 101, 40129 Bologna, Italy\\
$^{3}$Astrophysics Research Center of the Open University (ARCO), The Open University of Israel, P.O Box 808, Ra’anana 43537, Israel\\
$^{4}$Department of Astronomy, School of Science, The University of Tokyo, Bunkyo-ku, Tokyo 113-0033, Japan\\
$^{5}$Department of Astronomy and Astrophysics, The Pennsylvania State University, 525 Davey Lab, University Park, PA 16802, USA\\
}

   \date{Received September 30, 20XX}

% \abstract{}{}{}{}{}
% 5 {} token are mandatory
 
  \abstract
  % context heading (optional)
  % {} leave it empty if necessary  
   {}
  % aims heading (mandatory)
   {Recent observations have identified a new class of long-duration gamma-ray bursts (GRBs) likely produced by compact binary mergers. Using radio observations and lightcurve modelling, we investigate whether a magnetar remnant can power the long-lasting prompt emission observed in this merger-driven GRB sample. We derive constraints on the magnetars’ rotational energy and magnetic-field strength, and assess whether the magnetar central-engine model is viable for the observed events.} 
  % methods heading (mandatory)
   {We conducted 2.1 GHz radio observations of seven nearby merger-driven GRBs ($z<0.25$) using the Australia Telescope Compact Array (ATCA), obtaining sensitive upper limits on late-time radio emission 520–6900 days post-burst. We modelled the expected radio light curves from magnetar-energized ejecta interacting with the circumburst medium and compared them to our observations. We separately tested the magnetar hypothesis against the extended gamma-ray emission using a fallback-accreting magnetar model.}
  % results heading (mandatory)
   {No radio counterpart is detected in any of the observed GRBs, with $3\sigma$ flux density upper limits of $33$--$72\,\mu$Jy at 2.1\,GHz. Modelling the expected synchrotron emission from magnetar-energized ejecta interacting with the circumburst medium, we find that the radio non-detections remain compatible with energetic outflows in low-density environments. 
   The analysis of the prompt gamma-ray emission provides complementary constraints on the spin period and magnetic field of the magnetar engine. Once fallback accretion and jet baryon-loading are considered, the parameter space is narrowly confined.}
  % conclusions heading (optional), leave it empty if necessary  
  {}
  

   \maketitle
\nolinenumbers

%%%%%%%%%%%%%%%%%%%%%%%%%%%%%%%%%%%%%%%%%%%%%%%%%%%%%%%%%%%%%%
\section{Introduction}
Gamma-ray bursts (GRBs) are intense flashes of high-energy radiation, traditionally divided into two distinct classes based on their durations and spectral properties \citep{Kouveliotou1993,review_zhang_mesaros, review_piran, review_geherls,Kumar2015}, with a separation at approximately $T_{90} \approx 2$~s\footnote{$T_{90}$ is the time interval during which 90\% of the background-subtracted counts are detected}.  
Long-soft GRBs (lGRBs) are typically associated with the collapse of massive stars \citep{Woosley1993}, and are commonly found in star-forming regions \citep{Fruchter2006,Bloom2002}. Some of these events are accompanied by bright supernovae (SNe; \citealt{Woosley2006}), observable when the GRB occurs at relatively nearby distances. In contrast, short-hard GRBs (sGRBs) are widely believed to originate from binary neutron star (BNS) mergers or black hole-neutron star (BH-NS) mergers \citep{Paczynski1986,Eichler1989,Narayan1992,Ruffert1998,Kluzniak1998,Janka1999}, which are also a primary source of gravitational waves (GWs; \citealt{Abbott2017a,Abbott2017,Abbott2018d}). This scenario is further supported by the diversity of their host galaxy environments, ranging from star-forming to quiescent galaxies \citep{Gehrels2005,Fox2005,OConnor2022,Fong2022}. 
Most importantly, the connection between sGRBs and BNSs was definitively confirmed with the joint detection of GW170817 \citep{Abbott2017} and its electromagnetic counterpart, GRB~170817A \citep{Savchenko2017,Goldstein2017,Zhang2018}. 

On the other hand, a new class of GRBs exhibiting hybrid high-energy properties has emerged in recent years. Notable examples include GRB~060614 \citep{Gehrels2006,Jin2015}, GRB~211211A \citep{Troja2022,Rastinejad2022,JunYang2022} and GRB~230307A \citep{Yang2024,Levan2024}, which, based on their long durations, would typically be classified as lGRBs. However, the absence of bright SNe and the identification of kilonova\footnote{The thermal transient powered by radioactive decay in neutron-rich material ejected during a compact-object merger.} signatures instead point to a binary compact object merger origin. These findings challenge the traditional GRB classification scheme and suggest a more complex picture in which GRB duration alone is not a reliable indicator of progenitor type \citep{Bing2007,Virgili2011,Bromberg2011,Bromberg2013}.  

The duration of the gamma-ray emission probes the lifetime of the central engine. Constraining the nature of this engine is therefore key to physically understanding the new class of merger-driven GRBs, as different engines imply distinct energy reservoirs and characteristic timescales.
Currently, two competing models have been proposed for the central engines of GRBs: the accreting BH model and the magnetized NS model. In some scenarios, the NS promptly collapses into a BH within a timescale of $\lesssim 100$~ms, and the accretion of a thick remnant torus of mass $M_{\text{t}} \sim 10^{-2}\, \mathrm{M}_\odot$ powers the relativistic jet \citep{Narayan1992,Rezzolla2011}. The sole energy reservoir for the observed emission originates from the merger-formed accretion disk. However, in other cases, some NS mergers may instead result in the formation of a supramassive, highly magnetized, and rapidly rotating NS (a ms magnetar), with spin periods $P \sim 1$\,ms \citep[see e.g.][]{GiacomazzoPerna2013}. Such an outcome is possible for certain ranges of the progenitor NS masses, and the remnant can remain stable against gravitational collapse for an extended period \citep{Metzger2008, Bucciantini2012}. These magnetars may host magnetic fields as strong as $10^{15}$~G \citep{McKinney2006}, potentially amplified by shear instabilities or an $\alpha$-$\omega$ dynamo mechanism \citep{Duncan1992, Price2006}.

This magnetar-driven model has frequently been proposed as a possible explanation for certain distinctive $\gamma$-ray and X-ray features, such as the extended emission \citep{NorrisBonnell2006,Sun2025NSR} and X-ray plateaus \citep{Troja2007,Rowlinson2013, Dallosso2011,Dainotti2016}. The magnetar scenario offers an additional energy reservoir in the form of strong magnetic fields and rapid rotation. This can potentially account not only for the total energy of $\gtrsim 10^{52}$~erg observed in GRBs \citep{Duncan1992, Usov1992, ZhangMezoras2001, Dessart2008}, but also for their extended late-time X-ray activity.

Initially, the kinetic energy of the merger ejecta is enhanced by the continuous energy injection from the central magnetar. While only a small fraction of this energy is radiated away, the majority is deposited into the ejecta, accelerating it. Consequently, the resulting ejecta is expected to be faster than that produced by an accreting BH and may be re-energized at later times by the spin-down power of the NS. The total injected energy can exceed that of the dynamical ejecta by two to three orders of magnitude \citep{Nakar2011, MetzgerBower2014}. This also leads to an expectation of very bright magnetar-boosted KNe, the existence of which is strongly constrained by observations \citep{HaoWang2024}. As the ejecta propagate, it begins to decelerate at a characteristic radius $R_{\mathrm{dec}}$, when it sweeps up a mass from the surrounding medium comparable to its own. This occurs over a timescale $t_{\mathrm{dec}}$ \citep{Nakar2011}.

The interaction of sub-relativistic outflows with the ambient medium generates shocks that accelerate electrons. As these electrons spiral within magnetic fields, they emit synchrotron radiation, peaking around $t_\mathrm{dec}$ \citep{Nakar2011}. This process produces nearly isotropic non-thermal emission, predominantly in the radio (cm) band \citep{Nakar2011, Piran2013, Piro2013,Hotokezaka2015,Radice2018,Kathirgamaraju2019,Beniamini2021,Beniamini2017}. 
In the case of a black hole central engine, the radio brightness is governed solely by the kinetic energy of the dynamical ejecta. The expected signal is detectable only at relatively small distances ($\lesssim 300$~Mpc). 
However, if a long-lived magnetar remnant forms instead, its spin-down energy can be transferred to the surrounding ejecta, accelerating it to mildly relativistic velocities, producing radio emission orders of magnitude brighter and detectable at cosmological distances \citep{MetzgerBower2014}.
These radio signals may remain detectable for several months to years after the merger, providing a valuable observational window \citep{Nakar2011,Acharya2025,Rahaman2026}. In short, radio observations at $t \sim t_{\mathrm{dec}}$ are thus crucial for probing the total kinetic energy of the explosion, especially since early-time observations are largely insensitive to slower ejecta components. Therefore, long-term radio monitoring of merger-driven GRBs serves as a powerful tool for constraining the nature of the remnant and the dynamics of the merger.

Several previous studies have placed upper limits on radio emission associated with merger ejecta from sGRBs, providing crucial constraints on both the density of the surrounding medium and the rotational energy of potential magnetar remnants \citep{MetzgerBower2014, Horesh2016, Schroeder2020, Fong2016, Ricci2021, Ghosh2022}. In this paper, we select a sample of eight GRBs, all located at redshifts $z > 0.06$, where detection of the remnant is only feasible if the central magnetar significantly enhances the radio afterglow emission. 
We investigate the possibility of long-lived magnetar remnants through a comprehensive radio monitoring campaign using the \textit{Australia Telescope Compact Array} (\textit{ATCA}). Our study focuses on lGRBs without an associated SN. The sample selection criteria are described in Section~\ref{sec:sec2}, and details of our observations and analysis are presented in Section~\ref{sec:sec3}. Throughout the manuscript we adopt a standard $\Lambda$CDM cosmology \citep{Planck2020} with $H_0$\,$=$\,$67.4$\,km\,s$^{-1}$\,Mpc$^{-1}$, $\Omega_\textrm{m}$\,$=$\,$0.315$, and $\Omega_\Lambda$\,$=$\,$0.685$. 

\section{Sample Selection}
\label{sec:sec2}
In our study, we focus on a sample of lGRBs without an associated SN, such as GRB~060505, GRB~060614, GRB~211211A and GRB~230307A. A long-lived magnetar central engine has been invoked to account for their prolonged high-energy emission. These events were chosen for their precise localization ($\lesssim 1''$), relative proximity and the absence of an associated SN event, making them well-suited for testing the magnetar central engine model. The detailed properties of these GRBs are summarized in Table~\ref{tab:tab2}.

\subsection{GRB~230307A}

GRB~230307A was an exceptionally bright and long-duration event, with $T_{90} \approx 35$~s in the 10--1000~keV energy band characterized by a high gamma-ray fluence $\sim 3 \times 10^{-3}\,\mathrm{erg\,cm^{-2}}$ (10--1000~keV; \citealt{Yang2024}), yet a standard isotropic-equivalent energy of $E_{\gamma,\mathrm{iso}} \sim 3.2 \times 10^{52}\,\mathrm{erg}$ owing to its low redshift ($z = 0.0647$; \citealt{Yang2024,Levan2024}). 
Its prompt emission was followed by a pronounced X-ray plateau which is well described by a magnetar spin-down model with a magnetic field $B_p \sim 2.1 \times 10^{16}$~G and an initial spin period $P_0 \sim 3.3$~ms \citep{Sun2025NSR}. 
The most probable host, a spiral galaxy located at an offset of 30 arcsec (designated G1) at $z=0.0647$, is a low-mass, low star-formation-rate system ($M_\star \sim 2.4 \times 10^9\,M_\odot$, $\mathrm{SFR} \sim 0.2\,M_\odot\,\mathrm{yr^{-1}}$) (\citealt{Yang2024}).
Its environment and pronounced blue-to-red kilonova evolution are consistent with a compact object merger. 

\subsection{GRB~211227A}

GRB~211227A had a duration $T_{90}\approx84 \,\text{s}$ in the 15-350 keV energy band and fluence of $8.2 \times 10^{-6}\,\mathrm{erg\,cm^{-2}}$ (15--150~keV; \citealt{Lu2022}). This corresponds to an isotropic-equivalent energy of $E_{\gamma,\mathrm{iso}} \sim 1.1 \times 10^{51}\,\mathrm{erg}$ if placed at the redshift of its putative host galaxy, $z=0.228$ \citep{Lu2022}. The \textit{Swift}/BAT light curve exhibited a short, intense spike near $T_0$, followed by softer extended emission peaking around $T_0+11$~s, consistent with an sGRB with extended emission \citep{Tsvetkova2022GCN}. The putative host is a star-forming galaxy located $3.7^{\prime\prime}$ from the afterglow position and with a chance alignment probability of $P_{\mathrm{cc}} = 1.3\%$ \citep{Ferro2023}. It has a stellar mass of $M_\star \sim 3.4 \times 10^{10}\,M_\odot$ and a star formation rate of $\mathrm{SFR} \sim 0.84~M_\odot~\mathrm{yr^{-1}}$ \citep{Ferro2023}. Considering the host association is correct, despite extensive follow-up, no accompanying supernova was detected, and any associated kilonova would have been below the detection threshold.

\subsection{GRB~211211A}

GRB~211211A was a nearby long-duration GRB with $T_{90} \approx 51\,\text{s}$ \citep{Stamatikos2021GCN} in the 15-350 keV energy band with total fluence $\sim 3 \times 10^{-4}\,\mathrm{erg\,cm^{-2}}$ (15--150~keV; \citealt{Troja2022}), associated with a nearby galaxy, SDSS J140910.47+275320.8, at $z = 0.0763$ \citep{Troja2022,Rastinejad2022,JunYang2022} with probability of chance coincidence $P_{\rm cc} \sim 1.4\%$ \citep{Troja2022} and an isotropic-equivalent energy of $E_{\gamma,\mathrm{iso}} \sim 7.6 \times 10^{51}\,\mathrm{erg}$ (10--1000~keV; \citealt{JunYang2022}). The prompt emission displayed a bright initial hard spike of duration $\sim 10$~s, followed by softer extended emission starting after a 3~s low-level interval \citep{Troja2022,JunYang2022}. \citet{Troja2022} reported fluences of $\sim 3.7 \times 10^{-4}$ and $\sim 5 \times 10^{-5}\,\mathrm{erg\,cm^{-2}}$ for the main and extended-emission episodes, corresponding to $E_{\gamma,\mathrm{iso}} \sim 5 \times 10^{51}\,\mathrm{erg}$ and $\sim 7 \times10^{50}\,\mathrm{erg}$, respectively in the 10--1000~keV band. 
No accompanying supernova was detected despite the deep limits, ruling out a collapsar origin \citep{Troja2022,Rastinejad2022}. Instead, rapid optical/infrared follow-up revealed a luminous, rapidly evolving transient with early blue emission and fast reddening on timescales of $\lesssim 1$~day, consistent with an $r$-process kilonova \citep{JunYang2022,Rastinejad2022,Troja2022}.

\subsection{GRB~191019A}

GRB~191019A had a duration of $T_{90} \approx 64\,\text{s}$ 
in the 15--350~keV band, with an observed fluence of 10$^{-5}\,\mathrm{erg\,cm^{-2}}$ (15--150~keV; \citealt{Krimm2019GCN,Stratta2025}). Follow-up optical and near-infrared observations localized the transient to within $0.03^{\prime\prime}$ of the nucleus of an early-type, quiescent galaxy at $z = 0.248$ \citep{Levan2023}, corresponding to an isotropic-equivalent prompt energy of $E_{\gamma,\mathrm{iso}} \sim 1.7 \times 10^{51}\,\mathrm{erg}$ \citep{Krimm2019GCN, Levan2023}. The afterglow position is consistent with the galaxy nucleus, with a projected offset of $78 \pm 109$~pc \citep{Levan2023}. 
The old stellar population, negligible ongoing star formation, and deep limits on any accompanying supernova emission disfavor a massive-star origin \citep{Levan2023}. Instead, GRB~191019A has been interpreted as a possible compact object merger formed through dynamical interactions in the dense nuclear environment of its host \citep{Levan2023,Lazzati2023,Wang2024}. Late-time optical residuals at the GRB position have been reported and tentatively interpreted as consistent with a kilonova-like component \citep{Stratta2025}.  

\begin{table*}
\centering
\renewcommand{\arraystretch}{1.3}
\caption{Summary of observations and inferred ejecta/environment parameters for the merger-driven GRB sample. Column 2 lists the redshift ($z$); Column 3 gives $T_{90}$ in the 15--350\,keV band, in seconds, except for GRB~230307A, for which $T_{90}$ is reported in the 10--1000\,keV band; Column 4 notes the detection of a kilonova component; Column 5 lists the time since the burst ($T-T_0$) of our radio observations in days; Column 6 gives the 3$\sigma$ upper limit on the flux density at our observing frequency of 2.1\,GHz; Columns 7--9 list the ejecta mass, circumburst number density and projected host offset inferred from afterglow and kilonova modelling in the literature with references in Column 10. }
\label{tab:tab2}
\small
\setlength{\tabcolsep}{3.5pt}
\begin{tabular}{c c c c c c c c c >{\raggedright\arraybackslash}p{2.6cm}}
\toprule
\textbf{Target name} & \textbf{z} & \(\mathbf{T_{90}}\) & \textbf{Candidate KNe} & \(\mathbf{T-T_0}\) & \textbf{Flux} & \textbf{Ejecta mass} & \textbf{Density} & \textbf{Offset} & \textbf{References} \\
 & & [s] & & [d] & [\(\mu\)Jy] & [\(M_\odot\)] & [cm\(^{-3}\)] & [arcsec] & \\
\midrule
GRB 050911  & 0.16    & $16 \pm 2$      & -   & 6905 & $<33$ & $-$         & $<2.5\times10^{-1}$     & $-$               & \citet{Page2006,Berger2007} \\
GRB 060505  & 0.089   & $4 \pm 1$       & Yes & 6668 & $<72$ & $\sim 0.01$ & $>4\times10^{-5}$       & $4.30 \pm 0.03$   & \citet{Ofek2007,Xu2009,Jin2021} \\
GRB 060614  & 0.125   & $102 \pm 5$     & Yes & 6627 & $<60$ & $0.03-0.1$  & $\lesssim 0.04$         & $0.35 \pm 0.01$   & \citet{Gehrels2006,Yang2015} \\
GRB 130626A & Unknown & $0.16 \pm 0.03$ & -   & 4062 & $<57$ & $-$         & $\simeq 7\times10^{-2}$ & $4.1 \pm 0.2^{*}$ & \citet{Grandorf2021} \\
GRB 191019A & 0.248   & $64 \pm 4$      & Yes & 1751 & $<48$ & $\sim 0.04$ & $\lesssim 1$            & $0.02 \pm 0.02$   & \citet{Stratta2025,Levan2023} \\
GRB 211211A & 0.076   & $51 \pm 8$      & Yes & 971  & $<66$ & $0.01-0.1$  & $0.016\text{--}12$      & $5.35 \pm 0.03$   & \citet{Rastinejad2022,Troja2022} \\
GRB 211227A & 0.228   & $84 \pm 8$      & -   & 952  & $<33$ & $0.01-0.13$ & $<10^{-2}$              & $3.7 \pm 2.4$     & \citet{Ferro2023} \\
GRB 230307A & 0.065   & $34.6 \pm 0.6$  & Yes & 520  & $<66$ & $\sim 0.03$ & $\lesssim 0.03$         & $29.4 \pm 0.02$   & \citet{Sun2025NSR,Yang2024} \\
\bottomrule
\end{tabular}
\vspace{0.4em}

\textsuperscript{*}\, The reported separation corresponds to the distance between the VLA candidate source and the nearest NED infrared galaxy.
\end{table*}

\subsection{GRB~130626A}

GRB~130626A was a short duration burst with $T_{90} \approx 0.16\,\text{s}$ in the 15--350~keV band and a fluence of $5.2 \times 10^{-8}\,\mathrm{erg\,cm^{-2}}$ (15--150~keV; \citealt{Sakamoto2013GCN}). No X-ray and optical afterglow were confirmed, with the follow-up further limited by heavy Galactic extinction toward the burst, $E(B-V)=1.45$ \citep{Page2013GCN,Zheng2013GCN}. Consequently, no sub-arcsecond afterglow position, secure host association, or redshift could be obtained. However, late-time VLA observations by \citet{Grandorf2021}, taken at $\Delta t\simeq2203$--2468~d, identified an uncatalogued candidate radio source within the BAT localization region. Therefore, it was included in our sample to search for late-time radio flares. 

\subsection{GRB~060614}

GRB~060614 was a nearby burst at $z = 0.125$, with a long duration $T_{90} \approx 102\,\text{s}$ in the 15--350~keV band and a fluence of $2.2 \times 10^{-5}\,\mathrm{erg\,cm^{-2}}$ (15--150~keV; \citealt{Gehrels2006,Zhang2007}). Its prompt emission consisted of an initial short, hard episode lasting $\sim6$~s, followed by softer extended emission lasting $\sim100$~s \citep{JunYang2022}. The burst had a relatively low isotropic-equivalent gamma-ray energy $E_{\gamma,\mathrm{iso}} \sim 8.4 \times 10^{50}\,\mathrm{erg}$ \citep{Zhang2007}. No associated supernova was detected to deep limits \citep{Fynbo2006,Gal-Yam2006,DellaValle2006}. However, late-time HST optical observations showed a red F814W-band excess above the extrapolated afterglow decay, which was interpreted as
a possible kilonova component \citep{Yang2015,Jin2015}.

\subsection{GRB~060505}

GRB 060505 had $T_{90} \approx 4\,\text{s}$ in the energy range 15--350~keV band with a fluence $6.2 \times 10^{-7}\,\mathrm{erg\,cm^{-2}}$ (15--150~keV), which at its redshift $z=0.0894$
corresponds to $E_{\gamma,\mathrm{iso}} \sim 1.2\times10^{49}\,\mathrm{erg}$ \citep{Hullinger2006GCN,Ofek2007}. The host is a spiral galaxy, with the afterglow position located $\sim 6.5~\mathrm{kpc}$ from the galaxy center inside a bright H\,\textsc{ii} region hosting a very young stellar population ($\sim 6$ Myr), subsolar metallicity ($\sim 0.19$–$0.40~Z_\odot$) and with the site metallicity described as roughly $1/5\,Z_\odot$ \citep{Thone2008}. However, the burst's relatively short duration and lack of an associated SN led to the hypothesis
of a compact object merger progenitor. 

\subsection{GRB~050911}
 GRB 050911 had a duration of $T_{90} \approx 16\,\text{s}$ in the 15--350~keV band with a fluence $\sim 3 \times 10^{-7}\,\mathrm{erg\,cm^{-2}}$ (15--150~keV; \citealt{Page2006}). 
No X-ray afterglow was detected when \textit{Swift/XRT} began observations 4.6 hr after the trigger, with an upper limit of $F_{0.3-10 \mathrm{keV}}<1.7 \times 10^{-14}\,\mathrm{erg\,cm^{-2}\,s^{-1}}$ \citep{Page2006}. 
This led to the speculation of a possible compact object merger progenitor occurring in a low density environment \citep{Page2006}.  A galaxy-cluster (EDCC 493) at $z=0.1646$ was later proposed as putative host \citep{Berger2007}, corresponding to an
isotropic-equivalent gamma-ray energy of
$E_{\gamma,\mathrm{iso}} \sim 1.9 \times 10^{49}\,\mathrm{erg}$.

\section{Radio Observations and Data Reduction}
\label{sec:sec3}

ATCA observations were conducted between 2024 August 4 and August 10 using the 6A array configuration in the 16-cm band, centered at 2.1 GHz with a total bandwidth of 2 GHz (project C3532; PI: Troja). Target and phase calibrator sources were alternated every 20 minutes. The bandpass calibrator was PKS 1934–638. A full list of phase calibrators and observing sessions is presented in Table~\ref{table:tab1}. Data were processed using the Miriad software package \citep{Sault1995}. The data were split into single-source, single-band datasets. For GRB~050911, Antenna 3 exhibited an anomalously high system temperature (twice the nominal value), leading to elevated noise levels. Additionally, this observation was split into two segments: during the first half, Antenna 6 was offline but became operational in the second half. We calibrated each segment independently before combining them for imaging. The datasets were flagged for radio frequency interference (RFI) and antenna shadowing. Standard Miriad routines were then used for calibration and imaging. Cleaned and restored Stokes I maps were visually inspected using the DS9 visualization tool\footnote{\url{http://ds9.si.edu}} to search for transient radio emission at the GRB positions. The root-mean-square (rms) noise was estimated within source-free rectangular regions in each map. No significant radio emission was detected at the positions of any target GRBs. A summary of the observational results is provided in Table~\ref{tab:tab2}.

The clean components were used to refine the calibration in the visibility data through the phase self-calibration technique, which was run iteratively on the visibilities starting from an initial image model provided by the first clean components. 
Then the \texttt{SELFCAL} task is used for target self-calibration. The Clean Components (CC) obtained from initial imaging of the target serve as the starting point for the self-calibration model. Three iterations of self-calibration were done. More rounds could be done to further improve the model and enhance image quality for more complicated targets. We continue the invert, clean and selfcal procedure in a loop until the expected noise level is reached. We check the visibilities during this process to make sure the selected solution interval is adequate to remove all artifacts other than random noise. The procedure is over once another calibration cycle shows no improvements in terms of noise levels. 

\begin{table}
   % \centering
    \renewcommand{\arraystretch}{1.3}
    \begin{tabular}{lllll}
        \toprule
        \textbf{GRB} & \textbf{Date [UT]} & \textbf{Ph Cal} & \textbf{Array}\\
        \midrule
        230307A & 2024-08-08 21:05:40.9 & 0252-712  & 6A  \\
        211227A  & 2024-08-06 02:15:28.5 & 0837+012  & 6A  \\
        211211A  & 2024-08-08 06:41:07.2 & 1328+307  & 6A  \\
        191019A & 2024-08-04 16:10:45.7 & 2243-123  & 6A  \\
        130626A & 2024-08-09 10:58:29.8 & 1819-096  & 6A  \\
        060614 & 2024-08-05 17:12:01.8 & 2105-489  & 6A  \\
        060505 & 2024-08-06 15:10:38.4 & 2216-281  & 6A  \\
        050911 & 2024-08-07 18:58:43.5 & 0104-408  & 6A  \\
        \bottomrule
    \end{tabular}
    \caption{Phase Calibration and Array Configuration in GRB Observations}
    \label{table:tab1}
\end{table}

\section{The Radio Light Curve Modelling}
\label{sec:sec4}

\subsection{Physical Framework}
\label{sec:dynamics}

The late-time radio emission produced when merger ejecta decelerate in the circumburst medium provides a direct, nearly isotropic probe of the total kinetic energy deposited by the central engine \citep{Nakar2011,Bauswein:2013jpa}. If a long-lived magnetar forms after the merger, its spin-down luminosity is continuously injected into the surrounding ejecta, accelerating it to mildly relativistic velocities and boosting the resulting synchrotron signal by several orders of magnitude relative to the case of a black-hole remnant \citep{MetzgerBower2014}. Radio observations taken around the deceleration timescale $t_{\rm dec}$ are therefore uniquely sensitive to the presence of such a remnant.  We model this emission following the framework of \citet{Nakar2011, Hotokezaka2015, MargalitPiran2020, Ricci2021, MatsumotoPiran2021}, which we summarise here.

We treat the ejecta as a quasi-spherical shell of kinetic energy $E_{\rm ej}$ and mass $M_{\rm ej}$ expanding into a uniform medium of particle density $n$.  The initial ejecta velocity $\beta_0 \equiv v_0/c$ follows self-consistently
from energy conservation,
\begin{equation}
  E_{\rm ej} \simeq M_{\rm ej}\,c^2\,(\Gamma_0 - 1)\,,
  \label{eq:energy_conservation}
\end{equation}
where $\Gamma_0 = (1-\beta_0^2)^{-1/2}$ is the initial Lorentz factor. The ejecta expand freely until, at the deceleration time
\begin{equation}
t_{\text{dec}} = 5.4\,\mathrm{yr} \times (1 - \beta_0) \left[E_{\text{ej},51}^{1/3} \cdot n_0^{-1/3} \cdot \left( \frac{\beta_0}{0.3} \right)^{-5/3} \right] \,,
\label{eq:tdec}
\end{equation}
they have swept up a circumburst mass comparable to their own. Here we adopt the notation $Q_x = Q/10^x$, with all quantities in CGS units, as used above for $E_{\rm ej,51}$ and $n_0$. At this epoch the radio light curve peaks. By the timescales relevant here, the GRB's own relativistic afterglow has already faded below detectability; the ``jet'' that re-enters the discussion below is its decelerated, sub-relativistic remnant blast wave. Prior to $t_{\rm dec}$, this decelerated GRB jet sweeps the ISM ahead of the ejecta, temporarily suppressing the radio signal
until the ejecta overtakes the jet-driven Sedov--Taylor shell at the coalescence time \citep{MargalitPiran2020},
\begin{equation}
t_{\text{col}} = 0.2\,\mathrm{yr}\, E_{j,49}^{1/3} \, n_0^{-1/3} \left[ \frac{\xi (1 - \beta_0)}{\beta_0} \right]^{5/3} \,,
\label{eq:tcol}
\end{equation}
where \( E_{j,49} \) is the jet energy with $\xi = 1.17$ \citep{MargalitPiran2020,Ricci2021}. For all events in our sample, $t_{\rm obs} > t_{\rm col}$ within the density ranges constrained by afterglow and kilonova modelling
(Table~\ref{tab:tab2}), so jet quenching does not affect our conclusions \citep{MargalitPiran2020}.

The synchrotron flux follows the broken power-law spectrum of \citet{Sari1998}, modified for the sub-relativistic regime \citep[Eq.~17]{Ricci2021}.  The post-shock magnetic field $B = (8\pi\epsilon_B m_p n c^2 \beta^2 \Gamma^2)^{1/2}$ sets both the characteristic synchrotron frequency $\nu_m$ and the peak flux $F_{\nu_m}$.  At low velocities the minimum
electron Lorentz factor saturates at $\gamma_m = 2$ \citep[Eq.~18]{Ricci2021}, entering the deep-Newtonian regime below the critical velocity $\beta_{\rm DN} = (8m_e/\bar\epsilon_e m_p)^{1/2}$; see also
\citep{Granot2006,Sironi2013}. The full numerical implementation follows
\citet[Eqs.~1--10]{MatsumotoPiran2021}.

We adopt $p = 2.5$, $\bar{\epsilon}_e \equiv 4\epsilon_e(p-2)/(p-1) = 0.1$ \citep{Fong2016,Ricci2021},
and consider $\epsilon_B = 0.1$ and $0.01$. The ambient density spans $n = 10^{-6}$--$1\,\mathrm{cm^{-3}}$, encompassing values typical of rarefied intergalactic gas and the interstellar medium \citep{Ferriere2001,Meiksin2009,OConnor2020}. 

\begin{figure*}
    \centering
    \includegraphics[width=\linewidth]{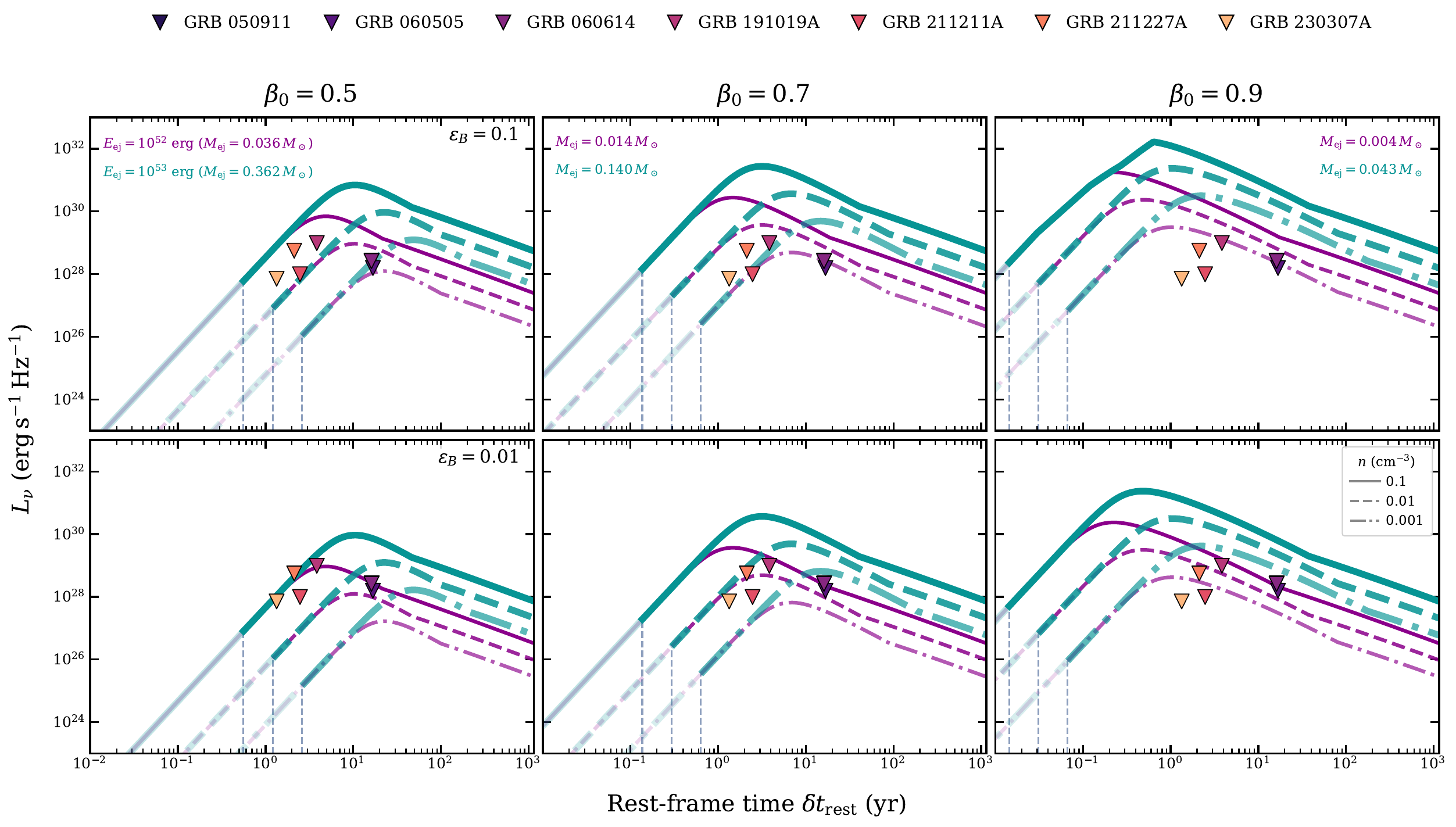}
    \caption{Spectral luminosity at 2.1\,GHz $L_\nu$ versus rest-frame time $\delta t_{\rm rest}$ where inverted triangles mark the observed $3\sigma$ upper limits. Rows correspond to fixed magnetic-field energy fraction $\epsilon_B = 0.1$ (top) and $0.01$ (bottom); columns correspond to $\beta_0 = 0.5$, $0.7$, $0.9$ excluding $\beta_0 = 0.3$ and $0.4$ as unphysical given the implied $M_{\rm ej}$ at $E_{\rm ej} = 10^{53}\,\mathrm{erg}$. In each panel, thin magenta curves show $E_{\rm ej} = 10^{52}\,\mathrm{erg}$ and thick teal curves show $E_{\rm ej} = 10^{53}\,\mathrm{erg}$, with the corresponding $M_{\rm ej}$ labelled in the top row. Curves of different line styles correspond to ISM densities $n = 10^{-1}$, $10^{-2}$ and $10^{-3}\,\mathrm{cm}^{-3}$. Microphysical parameters are $p = 2.5$ and $\bar{\epsilon}_e = 0.1$. Vertical dashed lines mark the coalescence time $t_{\rm col}$, computed assuming a fiducial jet kinetic energy $E_j = 10^{49}$\,erg; lighter portions of each curve indicate the quenched phase prior to $t_{\rm col}$.}
    \label{fig:lum_vs_time}
\end{figure*}

\begin{figure*}
    \centering
    \includegraphics[width=\linewidth]{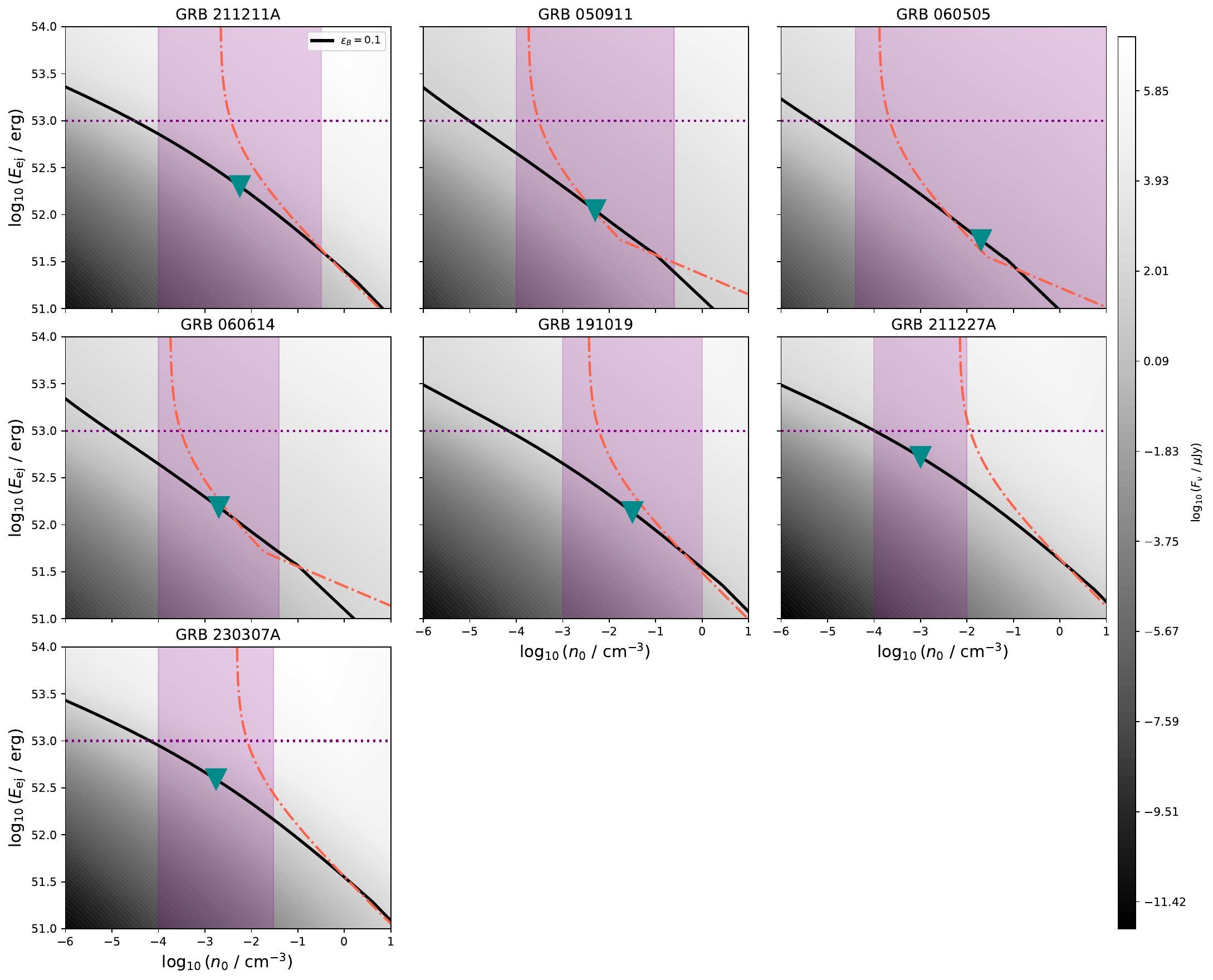}
    \caption{Ejecta energy ($E_{\mathrm{ej}}$) versus circumburst density ($n_0$) for seven GRBs, assuming an ejecta mass of $M_{\mathrm{ej}} = 0.05\,M_\odot$. Each panel shows the exclusion curve for the fiducial magnetic field energy fraction $\epsilon_B = 0.1$ (solid black line); regions above the curve are excluded. The gray-scale background shows the predicted flux density $\log_{10}(F_\nu / \mu\mathrm{Jy})$ at the rest-frame time of observation. The magenta hatched bands indicate the allowed range of $n_0$ inferred from afterglow and kilonova modelling (Table~\ref{tab:tab2}). Cyan inverted triangles mark the upper limit on $E_{\rm ej}$ at the median of the assumed density band. The orange dot-dashed contour shows the corresponding exclusion curve from the analytical model of \citep{Nakar2011}.}
    \label{fig:flux_contour}
\end{figure*}

\subsection{Model Parameters}
\label{sec:constraints}

We compute 2.1\,GHz model light curves for initial velocities $\beta_0 \in \{0.5, 0.7, 0.9\}$ chosen to span the range of ejecta masses expected from NS mergers, from $0.004$ to $0.36\,M_\odot$, covering the wind-driven component that a long-lived central engine can unbind \citep{Grossman2014, Radice2018Long-livedMergers, ciolfikalinani2020}. We choose kinetic energies $E_{\rm ej} \in \{10^{52}, 10^{53}\}$\,erg because $E_{\rm ej} = 10^{53}$~erg is the maximum rotational energy extractable from an indefinitely stable neutron star with baryonic mass of order $M_{\rm TOV}$, where spin-down proceeds to completion \citep{Metzger2015} and $E_{\rm ej} = 10^{52}$~erg, is a conservative lower bound on the energy extractable from a long-lived supramassive or stable remnant, applicable when the remnant mass is well below $M_{\rm TOV}$ or when collapse occurs before spin-down is complete \citep{Margalit2019}; the extractable rotational energy of such a remnant is typically of order a few $10^{52}$\,erg \citep{Duncan1992, Usov1992}. We do not explore energies below this range, like $E_{\rm ej} = 10^{51}$~erg, because for such low energies the remnant collapses promptly or survives only briefly as a hypermassive neutron star, releasing a small fraction of its rotational energy before collapse, forcing our model into an almost non-relativistic regime.

Figure~\ref{fig:lum_vs_time} compares the model light curves to our ATCA upper limits expressed as intrinsic spectral luminosities $L_\nu$. Looking at the light curves, at fixed $E_{\rm ej}$, increasing the density and the velocity shifts the peak to earlier times and higher luminosities, since a denser medium would decelerate the ejecta faster ($t_{\rm dec} \propto n^{-1/3}$, Equation \ref{eq:tdec}) while high $\beta_0$, observed for lower ejecta masses, will shorten the deceleration timescale ($t_{\rm dec} \propto \beta_0^{-5/3}$). Now increasing $E_{\rm ej}$ and $\epsilon_B$ increase the peak luminosity: the former through greater total energy in the ejecta, the latter through amplification of the post-shock magnetic field ($F_\nu \propto \epsilon_B^{(p+1)/4}$; \citealt{Sari1998}). Therefore, the parameter combinations most favourable for detection are high $E_{\rm ej}$, low $M_{\rm ej}$, high $n$, and high $\epsilon_B$. These are the conditions expected when a magnetar deposits a large fraction of rotational energy into a low-mass, fast-moving ejecta shell in a moderate-density environment \citep{Gao2013,MetzgerBower2014}. 

We adopt $\epsilon_B = 0.1$, the value chosen for modelling radio observations of Type~Ib/c supernovae \citep{ChevalierFransson2006} and later adopted by \citet{Nakar2011} for NS mergers, and $\epsilon_B = 0.01$ to show our sensitivity to fainter signals. Testing both lets us compare our constraints directly with previous radio searches for magnetar remnants \citep{MetzgerBower2014, Fong2016, Ricci2021, Schroeder2020, Ghosh2024}.

\begin{figure*}
    \centering
    \includegraphics[width=\linewidth]{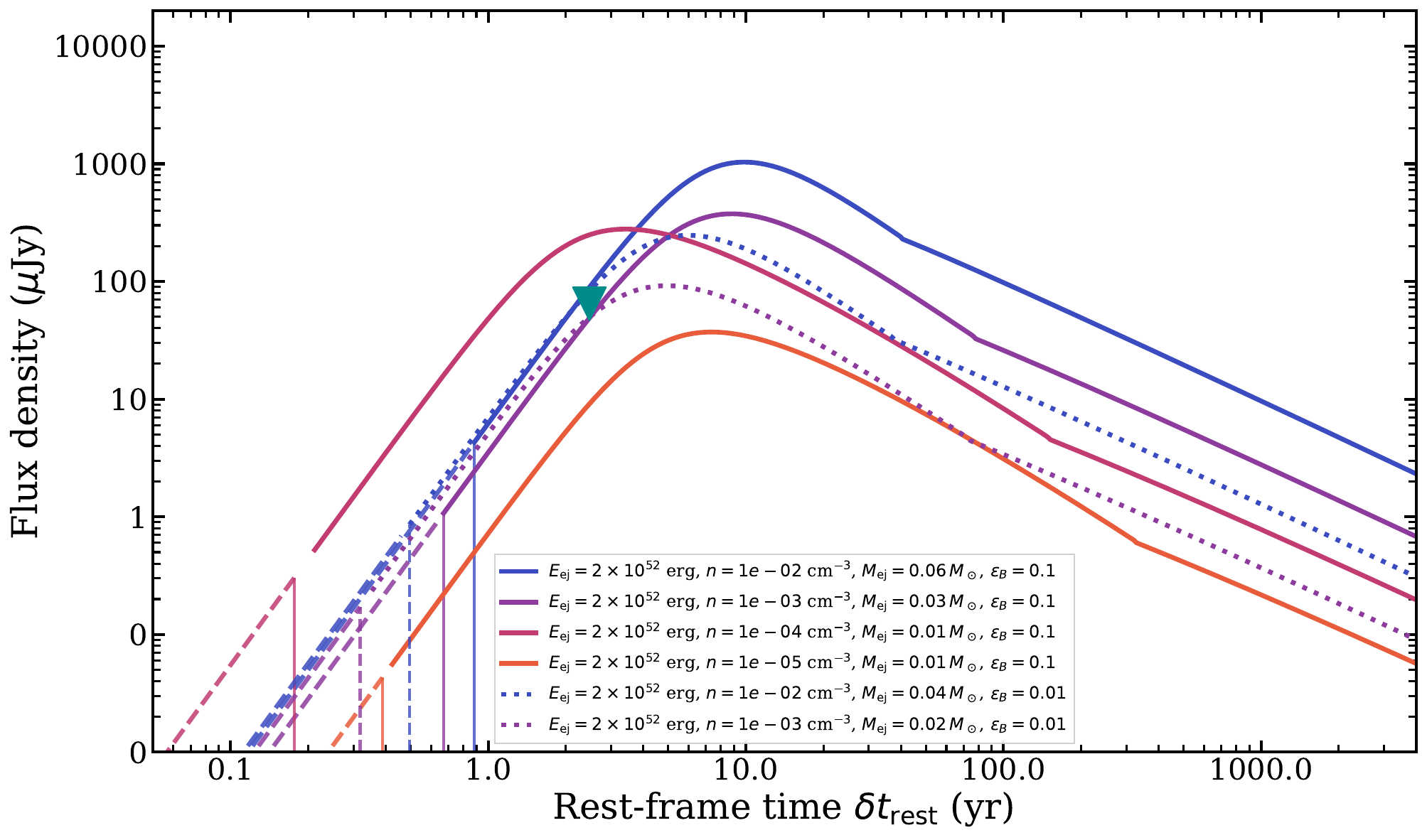}
    \caption{Model light curves of GRB~211211A ($z = 0.076$) computed assuming a rotational energy of $2\times10^{52}$\,erg (see Table~\ref{tab:magnetar_params}), for $\epsilon_B = 0.1$ and $0.01$, across a range of circumburst densities. The black inverted triangle marks the $3\sigma$ ATCA upper limit at 2.1\,GHz; curves whose predicted flux exceeds the upper limit at the observation epoch are excluded. Faded portions of each curve indicate the quenched phase before $t_{\rm col}$.}
    \label{fig:211211A}
\end{figure*}

\begin{figure*}
    \centering
    \includegraphics[width=\linewidth]{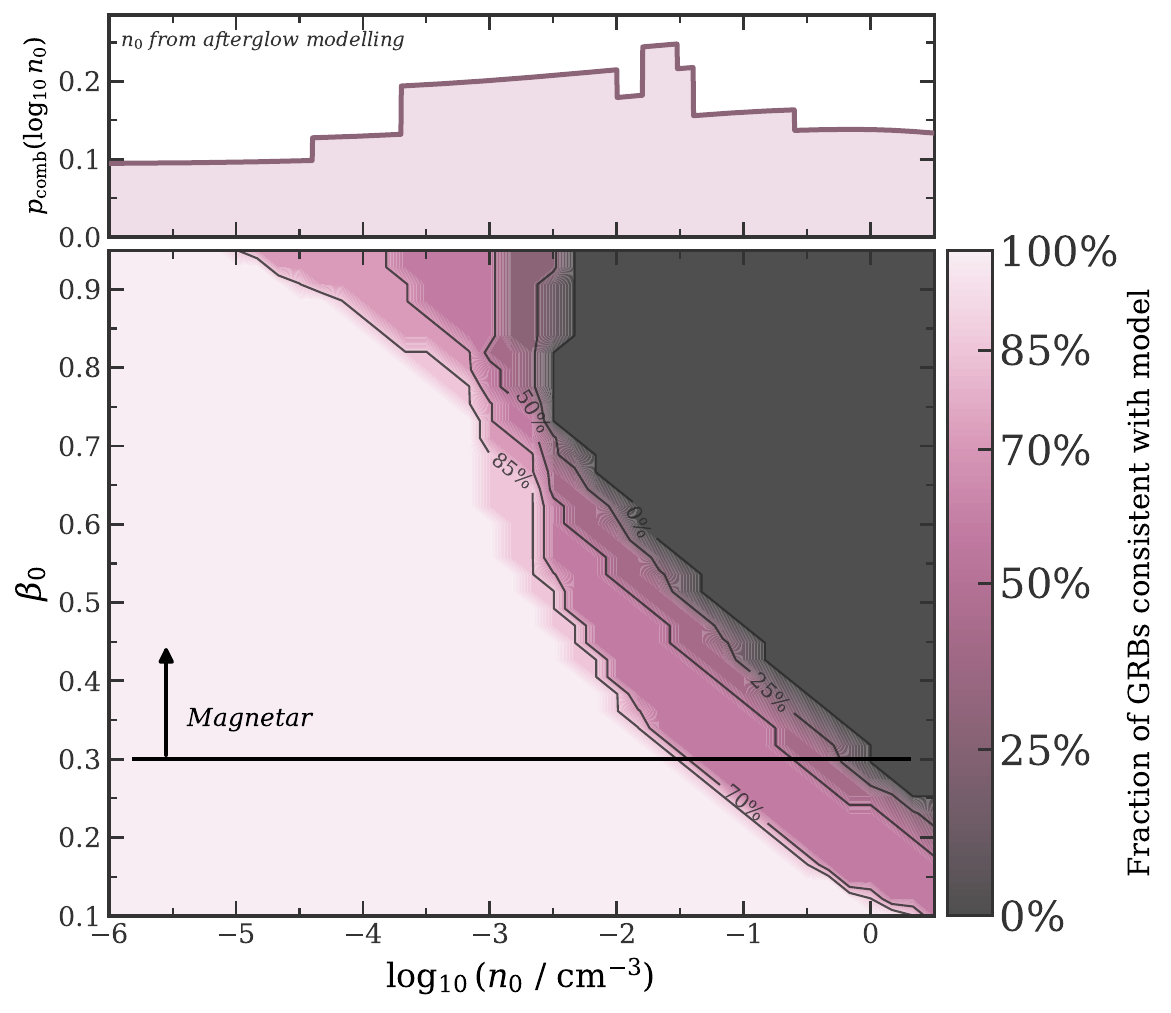}
    \caption{%
        \textit{Top panel:} Probability density profiles for the circumburst density \(n_0\) of each GRB in our sample, inferred from afterglow modelling (Table~\ref{tab:tab2}). The top panel shows the combined normalized distribution \(p_{\rm comb}(\log_{10} n_0)\), obtained by summing the individual GRB probability profiles and renormalizing the result. Upper and lower limits are represented as step-like probability distributions, and constrained ranges as flat-top profiles. 
        \textit{Bottom panel:} Fraction of the seven GRBs with known redshifts for which the predicted 2.1\,GHz flux at the observation epoch falls \emph{below} the measured $3\sigma$ upper limit, as a function of initial ejecta velocity $\beta_0$ and ambient density
        $\log_{10}(n_0)$, at fixed $E_{\rm ej} = 10^{52}$\,erg
        and $\epsilon_B = 0.1$.
        Light regions indicate that all events are consistent
        with the model; dark regions indicate widespread
        exclusion. The solid horizontal line at $\beta_0 = 0.3$ separates parameter space where the magnetar's contribution to ejecta acceleration is dominant ($\beta_0 > 0.3$; arrow) from the regime where sub-relativistic dynamical ejecta dominate.}
    \label{fig:combined}
\end{figure*}

\subsection{Comparison with Observations}
\label{sec:comparison}

Light curves in Sec.~\ref{sec:constraints} along with the individual ATCA upper limits disfavor energetic, fast-moving magnetar-driven ejecta ($\beta_0\gtrsim0.5$, $E_{\rm ej}=10^{52}$~erg) for circumburst densities typical of star-forming environments ($n_0\geq5\times10^{-2}\,{\rm cm}^{-3}$). 
The model remains consistent with the observations only 
for low densities 
($n_0\leq2\times10^{-3}\,{\rm cm}^{-3}$) or for slower, less energetic outflows ($\beta_0\lesssim0.2$). 

The model light curve rises while the ejecta is coasting and declines once deceleration becomes significant, near the characteristic timescale $t_{\rm dec}$. 
Therefore, measurements hold the greatest diagnostic power if the observations occur around this time (Figure~\ref{fig:lum_vs_time}). 
These are the cases of GRB~060614, GRB~060505, and GRB050911. 
For the most recent events, $t_{\rm obs}$ falls before $t_{\rm dec}$, catching the ejecta still on the rising branch.

For each burst, we compute the predicted 2.1~GHz flux at $t_{\rm obs}$ across a grid in $E_{\rm ej}$, at $M_{\rm ej}=0.05\,M_\odot$, $\epsilon_B=0.1$ and identify the energy at which this predicted flux equals the observed $3\sigma$ upper limit. The resulting exclusion contours in the $E_{\rm ej}$--$n_0$ plane are shown in Figure~\ref{fig:flux_contour}. It also compares our exclusion contours to the analytic approximation of \citet{Nakar2011} (orange dot-dashed curve). The two agree at moderate energies but diverge above
$E_{\rm ej}\gtrsim10^{52}$~erg; \citep{Nakar2011} state explicitly that their scaling relations assume $\Gamma_0-1\lesssim1$ and neglect relativistic corrections, whereas our implementation follows deep-Newtonian to mildly-relativistic synchrotron treatment of \citet{MatsumotoPiran2021}, which we find becomes progressively more important as the ejecta velocity and hence $E_{\rm ej}$ at fixed $M_{\rm ej}$, increases. 

We split the sample into three groups based on their inferred circumburst environment. The first group has GRB~211211A, GRB~211227A and GRB~230307A, which have large projected physical offsets from their host galaxies (Table~\ref{tab:tab2}), indicating that the binary merged far from any star-forming region, likely in a rarefied environment \citep{Troja2022,Yang2024}. This is independently supported, for all three bursts, by an anomalously low X-ray-to-gamma-ray flux ratio \citep{Troja2022}, a known signature of low circumburst densities \citep{OConnor2020}. These are also the events with the most extended prompt emission and the highest gamma-ray fluence in our sample, and represent the most promising magnetar candidates. As we discuss below, 
their non-detection do not disfavour the magnetar scenario.

For GRB~211227A and GRB~230307A only upper bounds on the density are available ($n\lesssim10^{-2}$ and $n\lesssim0.03$~cm$^{-3}$, respectively). Owing to their low density environment, a magnetar engine remains consistent with the observations across the entire velocity and energy range we consider, up to the fastest and most energetic outflow ($\beta_0=0.9$, $E_{\rm ej}=10^{53}$~erg). 

For GRB~211211A the allowed density range is broader. Independent estimates disagree by orders of magnitude, ranging from $n\approx10^{-4}$~cm$^{-3}$ inferred from afterglow deceleration \citep{Zhang2022} to $n \approx10^{-3}\,\mathrm{cm}^{-3}$ to $n\sim12$~cm$^{-3}$ from the jet-break fit \citep{Troja2022}. However, for a broad range of values within the allowed range, our upper limits can place meaningful constraints. At the best-fit density from \citep{JunYang2022} ($n\approx1.6\times10^{-3}$~cm$^{-3}$), the non-detection limits $E_{\rm ej}\lesssim3.1\times10^{52}$~erg comparable to the energy scale relevant to a magnetar engine.
For higher densities $n\gtrsim0.5$~cm$^{-3}$, non-detection firmly excludes the magnetar energy range (Figure~\ref{fig:211211A}), consistent with the constraint from \citet{Schroeder2025}.
A magnetar engine remains consistent only with the lower end of the possible density range, in agreement with the large offset of the burst. 

Moving to the second group, GRB~060614 and GRB~191019A have small projected offsets, placing them within their host galaxies, where densities are expected to be typical of the interstellar medium. Their observing baselines correspondingly probe closer to the light-curve peak, making these non-detections more constraining. For both events, the non-detections exclude the entire magnetar-relevant energy range ($E_{\rm ej}\lesssim10^{52}$~erg) toward the upper end of their allowed density range (Table~\ref{tab:tab2}), as for GRB~060614 at $n=0.04$~cm$^{-3}$, $t_{\rm dec}\sim10$~yr lies within the $\sim16$~yr rest-frame baseline, and for GRB~191019A at $n=1$~cm$^{-3}$, $t_{\rm dec}\sim3.5$~yr falls within the $3.84$~yr rest-frame baseline. Whereas at lower densities, $t_{\rm dec}$ will exceed $t_{\rm obs}$, leaving the ejecta still in their rising phase at $t_{\rm obs}$ and the non-detections not much informative.

The third group contains GRB~060505 and GRB~050911. For GRB~060505, the afterglow is poorly sampled, and only a lower bound on density is available ($n\gtrsim4\times10^{-5}$~cm$^{-3}$). However, its location within a H\,{\sc ii} region \citep{Thone2014}
favors much higher density values. Interestingly, our modelling shows that for $n\gtrsim4\times10^{-3}$~cm$^{-3}$ the entire magnetar-relevant range is ruled out by our limits. 

For GRB~050911, the radio non-detection excludes the relevant range of energies ($E_{\rm ej}\lesssim 10^{52}$~erg) for $n_0\gtrsim7\times10^{-3}$~cm$^{-3}$. Assuming that the burst lies within the galaxy cluster EDCC~493 at $z\approx0.16$ \citep{Berger2007}, we expect a circumburst density  $\approx10^{-4}$~cm$^{-3}$, typical of the intracluster medium. 
Therefore, there exists only a narrow range of densities compatible with a magnetar-driven explosion and the lack of a radio counterpart. 

To understand the collective constraining power of the sample, Figure~\ref{fig:combined} shows the fraction of events for which each combination of ejecta velocity ($\beta_0$) and ambient density ($n_0$) remains allowed. For fast ejecta expected from a magnetar-driven outflow, fewer than 25\% of events remain consistent with the model for ISM-like densities, $n_0\gtrsim3\times10^{-2}$~cm$^{-3}$. To remain consistent with the upper limits, fast-ejecta models imply low density environments. 

\begin{table*}
    \centering
    \renewcommand{\arraystretch}{1.3}
    \caption{Derived magnetar parameters for the GRB sample. The spin-down luminosity is calculated in the $1$--$10000$~keV energy band in the GRB rest frame. Rest-frame duration, $T_{100}$, is also provided. The initial spin period, $P_0$, refers to the spin of the magnetar formed immediately after the collapse. Both $P_0$ and the poloidal magnetic field strength, $B_p$, scale with the $\gamma$-ray efficiency factor, $\epsilon^{1/2}$, to account for uncertainties in the conversion from spin-down energy to $\gamma$-ray emission.}
    \label{tab:magnetar_params}
    \setlength{\tabcolsep}{6pt}
    \begin{tabular}{lcccc}
        \toprule
        \textbf{GRB} &
        \textbf{Luminosity} &
        $\bm{T_{100}}$ &
        $\bm{P_0}$ &
        $\bm{B_p}$ \\
        &
        ($10^{49}$~\si{erg.s^{-1}}) &
        (\si{s}) &
        ($\epsilon^{1/2}$~\si{ms}) &
        ($\epsilon^{1/2}\times10^{15}$~\si{G}) \\
        \midrule
        050911  & 0.66 & 15  & 15      & 170      \\
        060505  & 6.4  & 6.4 & $<7$    & $<130$   \\
        060614  & 1.9  & 161 & 2.6     & 9.4      \\
        191019A & 10   & 68  & 1.7     & 9.5      \\
        211211A & 19   & 112 & 1.0     & 4.2      \\
        211227A & 9.2  & 103 & 1.5     & 6.6      \\
        230307A & 65   & 42  & 1.0     & 6.0      \\
        \bottomrule
    \end{tabular}
\end{table*}

\begin{figure*}
    \centering
    \includegraphics[width=\linewidth]{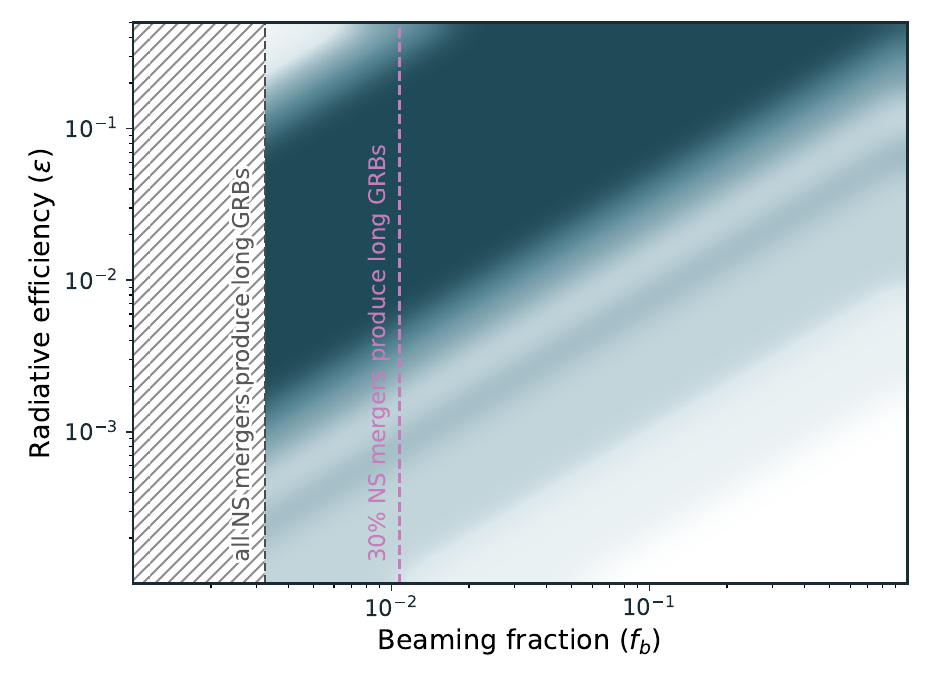}
    \caption{Allowed parameter space for the spin-down magnetar model. The shading indicates the number of bursts consistent with the model at each point in parameter space, with darker regions corresponding to more bursts. The hatched region at low $f_b$ is ruled out by the maximum NS merger rate consistent with GW observations, assuming all NS mergers produce long GRBs. The orange dashed line marks the same limit assuming only $\sim$30\% of NS mergers produce long GRB.}
    \label{fig:spindown_summary}
\end{figure*}

\section{Constraints from the Fallback Accreting Magnetar model}
\label{sec:sec5}

The late-time radio limits constrain the total kinetic energy deposited into the merger ejecta. 
However, a long-lived magnetar must not only supply the right total energy but must do so through a magnetically dominated relativistic jet that remains sufficiently magnetized and therefore baryon-clean throughout the entire observed gamma-ray emission episode. This places simultaneous constraints on the engine magnetic field, spin period and fallback accretion rate that are entirely independent of the circumburst density and the ejecta microphysics that enter the radio analysis.

We test the magnetar engine in two complementary steps, in addition to the radio constraints presented in Sect.~\ref{sec:constraints}. In Section~\ref{sec:spindown_params} we derive $P_0$ and $B_p$ directly from the observed average luminosity and duration under the standard magnetic dipole spin-down model \citep{ZhangMezoras2001}. In Section~\ref{sec:fallback_framework} we test whether any $(B_d, P_0)$ combination can simultaneously satisfy the energy and duration requirements of the observed gamma-ray emission when fallback accretion is self-consistently included \citep{MetzgerBeniaminiGiannios2018} which also accounts for jet baryon-loading. Together, these two diagnostics determine whether the magnetar hypothesis is viable for each burst and, if so, under what conditions. Throughout this section, we adopt the total GRB duration $T_{100}$ as a proxy for the engine lifetime. 

\subsection{Spin-Down Parameters from the Prompt and Extended Emission}
\label{sec:spindown_params}

Under magnetic dipole spin-down \citep{ZhangMezoras2001}, the observed average luminosity $L_0$ and duration $T_{100}$ determine the initial spin period $P_0$ and dipolar magnetic field strength $B_p$ of the remnant \citep[e.g.][]{Troja2007,Rowlinson2010,Lu2015}. The average luminosity was derived from the average flux in the 15--150~keV band obtained from the official \textit{Swift}/BAT online catalog,\footnote{https://swift.gsfc.nasa.gov/results/batgrbcat/} converted to the rest-frame 1--10000\,keV band (see Table~\ref{tab:magnetar_params}).

The initial poloidal magnetic field and spin period were then calculated using Equations~(6) and (8) from \citet{ZhangMezoras2001}. It is worth noting that GRB~060505 was identified through ground-based analysis. The available BAT event data span only 10\,s and the duration was estimated via FRED-profile fitting. Therefore, the measured duration ($\approx$6\,s) should be considered a lower limit, making the derived values of $B_p$ and $P_0$ upper limits. For GRB~230307A, detected by \textit{Fermi}/GBM but not seen by \textit{Swift}, we adopted the results reported by \citet{Dichiara2023}. Both $P_0$ and $B_p$ scale as $\epsilon^{1/2}$, where $\epsilon$ is the $\gamma$-ray radiative efficiency (see Table~\ref{tab:magnetar_params}, listing values at $\epsilon = 1$).

The inferred $P_0$ spans $1$--$3\,\mathrm{ms}$ and $B_p$ spans $4$--$10 \times 10^{15}$\,G. Both ranges are physically plausible for a newly formed millisecond magnetar. However, these estimates rely on the assumption of maximum efficiency $\epsilon = 1$ and isotropic emission. More generally, defining $f_b$ as the beaming factor, both parameters scale as $P_0,B_p\propto(\epsilon/f_b)^{1/2}$. Values lower than $\epsilon \approx 0.2$ would push the spin period close to the NS breaking limit \citep{Lattimer2007, LuJun2014}. Collimation alleviates this constraint by reducing the energy requirement although it correspondingly increases the magnetic field as well as the rate of events.

For GRB~050911 and GRB~060505, their low average luminosity, $L_0 \sim E_{\gamma,\mathrm{iso}}/T_{100}$, and short duration yield $B_p \sim 10^{17}$\,G, over an order of magnitude above the expected maximum field \citep[$\lesssim$\,a\,few\,$\times 10^{16}$\,G;][]{Duncan1992}. 
To reconcile these two bursts with the magnetar model, one must invoke either a low ($\epsilon<0.1$) radiative efficiency or a longer duration, missed by the limited sensitivity of the gamma-ray detectors. 

Figure~\ref{fig:spindown_summary} summarizes the joint constraints in $(f_b,\epsilon)$ space for all bursts. Each burst's allowed region forms a band bounded by the NS breakup limit ($P_0\gtrsim P_{\rm min}$) and the adopted maximum magnetic field \citep{Duncan1992}.
The event rate provides a lower bound on $f_b$. Adopting an intrinsic rate of $R_{\rm LGRB}\approx0.5\,f_b^{-1}\,{\rm Gpc^{-3}\,yr^{-1}}$ for merger-driven long GRBs (Yang et al. 2026, in prep.), consistency with the binary NS merger rate ($R_{\rm BNS}\lesssim155\,{\rm Gpc^{-3}\,yr^{-1}}$; \citealt[GWTC-5.0,][]{GWTC5}) requires $f_b\gtrsim0.003$. This limit assumes that every binary NS merger produces a long GRB. If only a fraction $\xi$ of mergers contributes to 
produce long GRBs, then the lower bound on $f_b$ increases to
$ f_b\gtrsim0.003\,\xi^{-1}$, excluding a larger portion of the parameter space.

\subsection{The Fallback-Accreting Magnetar Framework}
\label{sec:fallback_framework}

In addition to reproducing the observed luminosity and duration, 
the magnetar must sustain a relativistic outflow for the full duration $T_{100}$. 
In the fallback-accreting magnetar model \citep[e.g.][]{MetzgerBeniaminiGiannios2018}, the post-merger debris falls back onto the remnant on timescales of $\sim\!0.01$--$10$\,s with an accretion rate
\begin{equation}
    \dot{M}(t) = \frac{2}{3}\frac{M_\mathrm{fb}}{t_\mathrm{fb}}
    \left(1 + \frac{t}{t_\mathrm{fb}}\right)^{-5/3}\,
\end{equation}
where $M_{\rm fb}$ and $t_{\rm fb}$ are the fallback mass and fallback timescale, respectively, which is consistent with the late-time behavior expected from Keplerian debris streams \citep{MetzgerBeniaminiGiannios2018}. This accretion simultaneously sustains the magnetar spin-down and loads the jet base with baryons where the $\gamma$-ray production is governed by the jet magnetization,
\begin{equation}
    \sigma = \frac{\dot{E}_\mathrm{mag}}{\dot{M}_j c^2}\,
    \label{eq:sigma}
\end{equation}
where $\dot{E}_\mathrm{mag}$ is the electromagnetic spin-down power and $\dot{M}_j$ is the baryonic mass-loading rate at the jet base. In practice, this means the magnetar can only power $\gamma$-ray emission while its wind is neither too baryon-rich to accelerate relativistically ($\sigma < 100$) nor so baryon-free that the jet becomes radiatively inefficient ($\sigma > 3000$) \citep{Beniamini:2017ilu, MetzgerBeniaminiGiannios2018}. 
The baryonic loading $\dot{M}_j$ receives contributions from a neutrino-cooling wind (dominant at early times, $t \lesssim 20$~s) and an accretion-powered component that scales with the fallback rate; since $\sigma$ decreases with increasing $\dot{M}_j$, larger fallback mass suppresses $\sigma$, narrowing or closing the viable parameter space.

For a given combination of $B_d$ and $P_0$, the fallback-accreting model is considered viable if it simultaneously satisfies two conditions.
\begin{enumerate}
    \item \textit{Energy condition}: the total electromagnetic energy released while $\sigma \in [100, 3000]$,
    \begin{equation}
        E_\mathrm{avail} = \int_{\sigma \in [100,\,3000]} \dot{E}_\mathrm{mag}\,dt\,,
    \end{equation}
    must exceed the beaming-corrected energy required,
    \begin{equation}
        E_\mathrm{req} = \frac{f_b\, E_{\gamma,\mathrm{iso}}}{\epsilon}\,,
    \end{equation}
    with $f_b = 1 - \cos\theta_j$ the beaming fraction representing the solid angle covered by the jet and $\epsilon = 0.2$ is the assumed $\gamma$-ray efficiency \citep{Frail2001,Fan2006,Nava2014,Beniamini2016}.
    \item \textit{Duration condition}: the jet remains in the allowed $\sigma$ window for the full $T_{100}$ duration.
\end{enumerate}

Both these conditions depend on the NS mass. The mass sets the stellar structure entering $\dot{E}_\mathrm{mag}$ and, for a supramassive remnant, it also fixes the collapse period entering into the duration condition. 
Here we adopt $M_{\rm TOV}=2.1\,M_\odot$, whereas the case of a supramassive NS is explored in Section~\ref{sec:remnant_mass}. The true post-merger remnant mass is poorly constrained and likely lies in the range $\sim 2$--$2.5\,M_\odot$ \citep{Beniamini2021}. We adopt the conservative baseline $M_{\rm ns0}=2.05\,M_\odot$, at which the remnant is stable, and explore the impact of the full supramassive range in Section~\ref{sec:remnant_mass}. We scan the magnetar parameter space over $B_d \in [10^{14}, 10^{16}]$~G and $P_0 \in [0.6, 10]$~ms, where the lower edge, $P_{\rm shed,min}=0.6$\,ms, is the minimum period allowed by the mass-shedding (break-up) limit \citep{Breu2016}. The fallback mass is treated as a free parameter spanning the range expected from NS--NS merger simulations \citep{MetzgerBeniaminiGiannios2018}:
\begin{itemize}
    \item $M_\mathrm{fb} \in \{0.001, 0.01, 0.05, 0.1\}\,M_\odot$ at 
    $t_\mathrm{fb} = 0.1$~s,
    \item $M_\mathrm{fb} \in \{0.001, 0.01\}\,M_\odot$ at 
    $t_\mathrm{fb} = 0.01$~s,
    \item $M_\mathrm{fb} = 0.001\,M_\odot$ at $t_\mathrm{fb} = 10$~s.
\end{itemize}
          
\subsubsection{Fallback Viability Across the Sample}
\label{sec:fallback_results}

The fallback-accreting magnetar model is disfavored for the five bursts energetic enough to test it: GRB~060614, GRB~211211A, and GRB~211227A admit no viable solution under any fallback scenario tested, while GRB~191019A and GRB~230307A admit solutions only within a narrow range of extreme engine parameters. 
By contrast, GRB~050911 and GRB~060505 are too faint to test the model at all, and are therefore neither supported nor disfavoured by this analysis (see Table~\ref{tab:fallback_viability}).

This result derives from Equation~\eqref{eq:sigma}: the jet magnetization $\sigma$ falls as baryon loading rises and vice versa. A high fallback mass loads the jet with too many baryons, driving $\sigma$ below its lower limit. A low fallback mass supplies too few baryons to load the jet, so $\sigma$ remains high and exceeds its upper limit before $T_{100}$ is reached. Therefore, a viable $(B_d,P_0)$ solution requires an intermediate $M_{\rm fb}$ that keeps $\sigma$ within $[100,3000]$ for the full GRB duration.

\begin{table*}
\centering
\renewcommand{\arraystretch}{1.4}
\caption{
Constraints from the fallback-accreting magnetar model on the $\gamma$-ray emission of each GRB at $t_\mathrm{fb} = 0.1$~s. Columns list the beaming-corrected energy requirement $E_\mathrm{req} = f_b\,E_{\gamma,\mathrm{iso}}/\epsilon$ (with $\epsilon = 0.2$), followed by the allowed ranges of dipole field $B_d$ and initial spin period $P_0$ for four values of the fallback mass $M_\mathrm{fb}$. GRBs are grouped by their viability as:\\[6pt]$^a$\,\textit{Unconstrained}: energy requirement is so modest that virtually any magnetar configuration satisfies both conditions,$E_{\gamma,\mathrm{iso}} \lesssim 10^{49}$~erg, rendering these events uninformative as model tests.\\
$^b$\,\textit{Excluded} (``---''): no $(B_d,\,P_0)$ combination satisfies the energy ($E_\mathrm{avail} \geq E_\mathrm{req}$) and duration condition
($\sigma$-window duration $\geq T_{100}$) simultaneously for any fallback scenario explored.\\
$^c$\,\textit{Marginal}: solutions exist only at extreme parameters ($P_0 \sim 0.6$~ms, $B_d \sim 4$--$10 \times 10^{15}$~G) and the allowed region vanishes or shrinks further as $M_\mathrm{fb}$ increases.}
\label{tab:fallback_viability}
\setlength{\tabcolsep}{4.5pt}
\begin{tabular}{l r  r r  r r  r r  r r}
\toprule
& &
\multicolumn{2}{c}{$M_\mathrm{fb} = 0.001$} &
\multicolumn{2}{c}{$M_\mathrm{fb} = 0.01$} &
\multicolumn{2}{c}{$M_\mathrm{fb} = 0.05$} &
\multicolumn{2}{c}{$M_\mathrm{fb} = 0.1$} \\
\cmidrule(lr){3-4}\cmidrule(lr){5-6}\cmidrule(lr){7-8}\cmidrule(lr){9-10}
\textbf{GRB} &
\bh{$E_\mathrm{req}$} &
\bh{$B_d^{\phantom{0}}$} & \bh{$P_0$} &
\bh{$B_d^{\phantom{0}}$} & \bh{$P_0$} &
\bh{$B_d^{\phantom{0}}$} & \bh{$P_0$} &
\bh{$B_d^{\phantom{0}}$} & \bh{$P_0$} \\
&
(erg) &
 &  &
($10^{15}$~G) & (ms) &
 &  &
 &  \\
\midrule
\rowcolor{gray!8}
\multicolumn{10}{c}{\textit{Unconstrained:} $E_{\gamma,\mathrm{iso}} \lesssim 10^{49}$~erg} \\
\midrule
\rowcolor{gray!8}
050911  & $5.1 \times 10^{47}$ & all & all & all & all & $0.1$--$7.9$ & all & all & $0.6$--$4.1$  \\
\rowcolor{gray!8}
060505  & $3.0 \times 10^{47}$ & all & all & all & all & $0.1$--$7.9$ & all & all & $0.6$--$4.1$  \\
\midrule
\rowcolor{gray!8}
\multicolumn{10}{c}{\textit{Marginal:} Solutions at extreme parameters only} \\
\midrule
\rowcolor{gray!8}
191019A & $4.2 \times 10^{49}$  & --- & --- & $10.0$ & $0.60$ & $6.2$ & $0.60$ & $4.8$ & $0.60$--$1.09$  \\
\rowcolor{gray!8}
230307A & $2.8 \times 10^{50}$  & --- & --- & $6.2$--$10$ & $0.60$--$1.09$ & $3.8$--$4.8$ & $0.60$--$1.09$ & $3.8$--$4.8$ & $0.60$  \\
\midrule
\multicolumn{10}{c}{\textit{Excluded:} No viable $(B_d, P_0)$ for any $M_\mathrm{fb}$} \\
\midrule
060614  & $4.1 \times 10^{49}$  & --- & --- & --- & --- & --- & --- & --- & ---  \\
211211A & $1.9 \times 10^{50}$  & --- & --- & --- & --- & --- & --- & --- & ---  \\
211227A & $2.7 \times 10^{49}$  & --- & --- & --- & --- & --- & --- & --- & ---  \\
\bottomrule
\end{tabular}
\end{table*}
\textit{Excluded:} GRB~060614, GRB~211211A, and GRB~211227A admit no viable $(B_d, P_0)$ combination under any fallback scenario explored. Their long durations, $T_{100}\gtrsim100$~s, require the jet magnetisation to remain within the allowed $\sigma$ interval for an extended period.  This favours gradual spin-down, associated with a longer initial period or a weaker dipole field. Their large isotropic energies ($E_{\gamma,\rm iso}\sim10^{51}$--$10^{52}$\,erg) impose the opposite requirement: only a fast-spinning, strongly magnetised remnant can supply the required spin-down energy reservoir.
These two conditions cannot be satisfied simultaneously across the range of $M_{\rm fb}$ explored. 

\textit{Marginal:} For GRB~191019A and GRB~230307A, no solution is found at low fallback masses ($M_{\rm fb}=0.001\,M_\odot$). In this regime, the baryon loading is too weak to keep $\sigma$ below its upper limit for the required duration. At $M_{\rm fb}\gtrsim0.01\,M_\odot$, the higher baryon loading keeps $\sigma$ within range for longer. This opens a narrow region of allowed parameters, with periods close to the mass-shedding spin limit ($P_0\approx0.6$\,ms) and high magnetic field,  $B_d\sim10^{16}$\,G. 
As $M_{\rm fb}$ increases further, the allowed dipole field 
decreases down to $B_d\sim4$--$5\times10^{15}$\,G. Thus, these bursts are viable only within a highly fine-tuned region of the parameter space. 

Extending the fallback timescale to $t_{\rm fb}=10$~s can alleviate this tension by sustaining baryon loading to later times and thereby prolonging the interval over which $\sigma$ remains suitable for gamma-ray production. 
However, such a long fallback timescale lies well outside the range usually expected from NS--NS merger simulations \citep[$t_{\rm fb}\lesssim1$~s;][]{MetzgerBeniaminiGiannios2018}.

\textit{Unconstrained:} GRB~050911 and GRB~060505 have much lower isotropic-equivalent energies, $E_{\gamma,\mathrm{iso}}\sim10^{49}$~erg. Their required
engine energies are consequently small enough that the energy condition is satisfied across most configurations that also meet the duration requirement. 
Their low energy requirements do not provide a meaningful test of the fallback accreting magnetar model.

\begin{figure*}
\centering
\includegraphics[width=\textwidth]{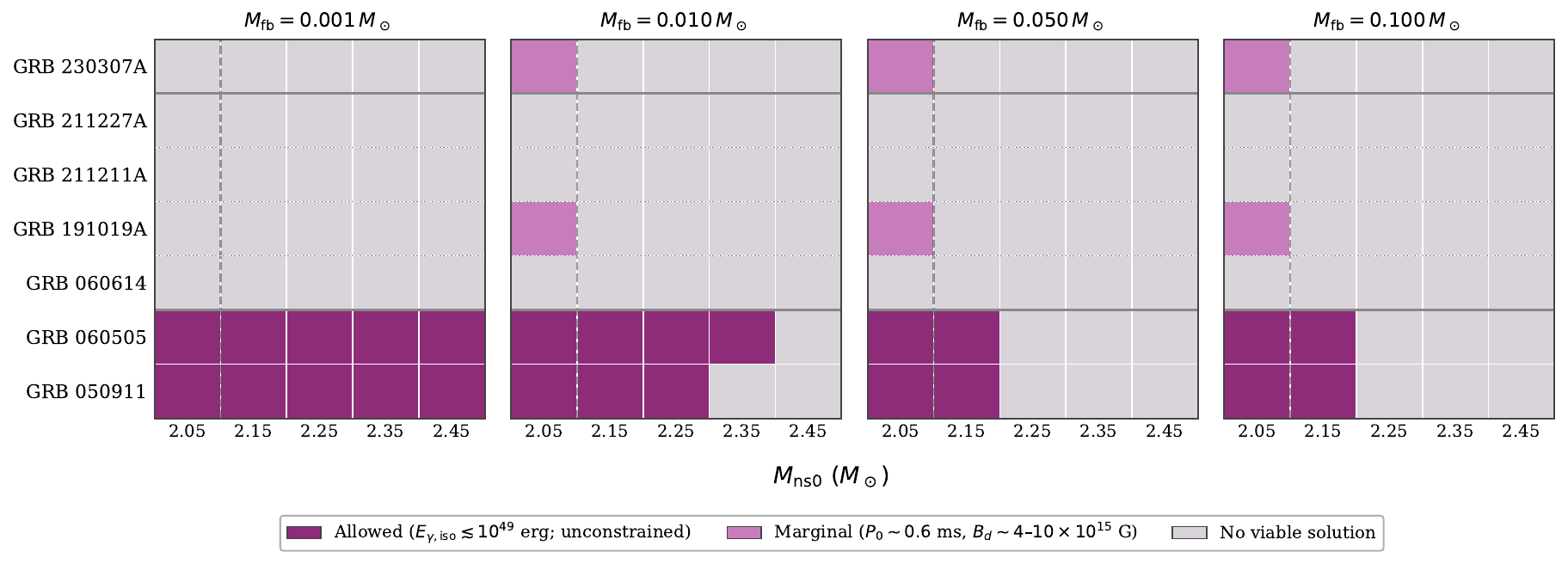}
\caption{Viability of the fallback-accreting magnetar model as a function of fallback mass $M_{\rm fb}$ and initial remnant mass $M_{\rm ns0}$, evaluated at $t_{\rm fb} = 0.1$\,s. Dark-purple cells denote an allowed $(B_d, P_0)$ region; light-purple cells denote marginal solutions requiring near-breakup spin periods ($P_0 \approx 0.6$\,ms) and extreme dipole fields ($B_d \sim 4$--$10 \times 10^{15}$\,G); light grey cells denote no viable solution. GRBs are grouped by their viability at the baseline mass $M_{\rm ns0} = 2.05\,M_\odot$: \textit{excluded} (no $(B_d,\,P_0)$ combination satisfies both the energy and duration conditions for any fallback scenario explored);
\textit{marginal} (solutions exist only at extreme parameters); and \textit{unconstrained} ($E_{\gamma,\rm iso} \lesssim 10^{49}$\,erg, too faint to test the model). The allowed parameter space shrinks progressively as $M_{\rm ns0}$ increases above $M_{\rm TOV} = 2.1\,M_\odot$, and the model is fully excluded for all seven constrained bursts at $M_{\rm ns0} = 2.45\,M_\odot$ with $M_{\rm fb} \geq 0.01\,M_\odot$. }
\label{fig:remnant_mass_viability}
\end{figure*}

\subsubsection{Dependence on the Remnant Mass}
\label{sec:remnant_mass}

A supramassive remnant ($M_{\rm ns0}>M_{\rm TOV}$) is formed in an estimated $45$--$90\%$ of binary neutron star mergers depending on the equation of state \citep{Beniamini2021}. Therefore, we test how the conclusions of Section~\ref{sec:fallback_results} hold up across the supramassive range by repeating the analysis for $M_{\rm ns0} \in \{2.05, 2.15, 2.25, 2.35, 2.45\}\,M_\odot$. 

Increasing the remnant mass imposes progressively more restrictive conditions. The physical reason is the growing requirement for rotational support. A supramassive remnant is held up by rotation alone. As it spins down, centrifugal support weakens until it can no longer support the remnant's mass, and the star collapses to a black hole \citep{Breu2016}. An initially supported remnant must therefore satisfy $P_{\rm shed,min}\leq P_0<P_{\rm col}$,
where $P_{\rm shed,min}$ is the minimum allowed period and $P_{\rm col}$ is the critical collapse period: 
\begin{equation}
    P_{\rm col} = \frac{P_{\rm shed,min}}{\sqrt{(M_{\rm ns0}/M_{\rm
    TOV} - 1)/0.2}}\,.
    \label{eq:Pcol}
\end{equation}

The maximum allowed $P_{\rm col}$ shrinks as the remnant mass increases. This further narrows the range of $P_0$ for which the remnant survives long enough to power the extended emission. For instance, $P_{\rm col}=1.74, 1.00, 0.78, 0.66$\,ms at $M_{\rm ns0}=2.15, 2.25, 2.35, 2.45\,M_\odot$, respectively. 
By $M_{\rm ns0}=2.35$--$2.45\,M_\odot$, the surviving window between break-up and collapse has shrunk to just $0.18$--$0.06$\,ms (Figure~\ref{fig:remnant_mass_viability}). Above this mass, the remnant leaves the rigidly-rotating supramassive regime. It becomes hypermassive, supported by differential rotation rather than rigid spin, and collapses on a dynamical, sub-second timescale too short to power any burst's extended emission, regardless of $(B_d,P_0)$.

\section{Summary and Conclusions}
\label{sec:conclusions}
\label{sec:discussion}

We presented late-time 2.1\,GHz ATCA observations of nearby ($z<0.25$) long-duration GRBs without an associated supernova, obtained 520--6900 days after the bursts. No radio counterparts were detected. We combined these observations with an analysis of the extended prompt emission to test the long-lived magnetar engine: the radio limits constrain the energy transferred to the merger ejecta, while the prompt emission analysis tests whether the engine can sustain the required energy output and duration. This extends previous radio searches for magnetar remnants in short GRBs \citep{MetzgerBower2014,Fong2016,Ricci2021,Ghosh2022} to the long-duration population.

The radio constraints depend strongly on the circumburst density. For GRB~060614 and GRB~191019A, the non-detections limit the ejecta energy to approximately $10^{52}$~erg at the upper end of their allowed density ranges. For GRB~211211A, the limit reaches $E_{\rm ej}\lesssim5\times10^{51}$~erg at $n\gtrsim0.5$~cm$^{-3}$. These bounds disfavor the substantial energy injection expected from a rapidly rotating, long-lived magnetar if the explosions occurred in an ISM-like environment. 
At lower densities, however, energetic ejecta remain consistent with our observations as their radio emission is still rising. This limits the diagnostic power of the observations, particularly for GRB~211227A and GRB~230307A, whose large galactocentric offsets favor low-density ($n\lesssim0.001$~cm$^{-3}$) environments.

Simple spin-down estimates \citep{ZhangMezoras2001} give plausible magnetar parameters for most of our sample, with $P_0\sim1$--$3$~ms and $B_p\sim(4$--$10)\times10^{15}$~G under isotropic emission. However, these estimates depend on the assumed radiative efficiency and beaming fraction, with $P_0,B_p\propto(\epsilon/f_b)^{1/2}$ (Section~\ref{sec:spindown_params}). Lower efficiency demands faster rotation and can push the shortest inferred periods below break-up. Collimation relaxes this spin constraint but increases the required magnetic field and the rate of events.  
Beaming therefore cannot be increased arbitrarily to compensate for low radiative efficiency: the allowed $(\epsilon,f_b)$ combinations must satisfy both the spin-down constraints and the GW rate of binary NS mergers. To explain the brightest events, such as GRB211211A and GRB230307A, the magnetar must convert its rotational energy into gamma-rays efficiently ($\epsilon \gtrsim$1\%). 

Our observations provide a stringent test for the fallback-accreting magnetar model \citep{MetzgerBeniaminiGiannios2018}. Within this framework, GRB~060614, GRB~211211A, and GRB~211227A admit no combination of magnetic field and initial spin period that satisfies both the energy and duration requirements. GRB~191019A and GRB~230307A remain marginal: solutions require rotation near break-up, $B_d\gtrsim4\times10^{15}$~G, and $M_{\rm fb}\geq0.01\,M_\odot$. They exist only at the lowest remnant mass tested, $M_{\rm ns0}=2.05\,M_\odot$, and disappear already at $2.15\,M_\odot$. GRB~050911 and GRB~060505 are less informative. Their low energies, $E_{\gamma,\rm iso}\sim10^{49}$~erg, impose only weak requirements on the fallback engine. However, the allowed range of solutions contracts as the remnant mass increases, and at $M_{\rm ns0}=2.45\,M_\odot$, solutions survive only for the lowest fallback mass explored, $M_{\rm fb}=0.001\,M_\odot$. 

Taken together, these results restrict the conditions under which a long-lived magnetar can explain the sample. The radio observations constrain substantial energy injection where the density permits a detectable signal, consistent with the tension identified by \citet{Beniamini2021}, while the spin-down analysis restricts the efficiency and beaming combinations compatible with the physical magnetar parameters and the binary NS merger rate. 

The fallback calculation requires the magnetar engine to reproduce both the burst energy and duration. This excludes or narrowly confines the allowed parameters for five events, although the exclusions remain dependent on the adopted emission and accretion prescriptions. 

Prompt or delayed black-hole formation, followed by fallback-powered extended emission, remains a viable alternative \citep{Rosswog2007,Kisaka2015,Musolino2024}. A magnetar could also evade the radio limits if much of its rotational energy escaped through gravitational waves \citep{DallOsso2015}, thus reducing the energy deposited into the quasi-spherical ejecta. Our analysis does not uniquely identify the central engine, but it shows that prolonged emission alone provides insufficient evidence for a long-lived magnetar. Later and deeper radio observations, together with tighter density estimates, will test the energetic remnants that remain allowed.

\begin{acknowledgements}
This work is supported by the European Research Council through the Consolidator grant BHianca (grant agreement ID~101002761). S.D. was supported by NASA under award number 80NSSC26K0666.
\end{acknowledgements}

%%%%%%%%%%%%%%%%%%%%%%%%%%%%%%%%%%%%%%%%%%%%%%%%%%%%%%%%%%%%%%
% WARNING
% Please note that we have included the references below in
% order to compile the document, but we ask you to:
%
% - use BibTeX with the regular commands:
%   \bibliographystyle{aa} % style aa.bst
%   \bibliography{Yourfile} % your references Yourfile.bib
% - join the .bib files when you upload your source files
%%%%%%%%%%%%%%%%%%%%%%%%%%%%%%%%%%%%%%%%%%%%%%%%%%%%%%%%%%%%%%

\bibliographystyle{bibtex/aa}  % A&A bibliography style
\bibliography{main}     % main.bib file name (without .bib extension)

% %%%%%%%%%%%%%%%%%%%%%%%%%%%%%%%%%%%%%%%%%%%%%%%%%%%%%%%%%%%%%%
% Example below of non-structurated natbib references  
% To use the v8.3 macros with this form of composition of bibliography,
% the option "bibyear" should be added to the command line
% "\documentclass[bibyear]{aa}".
% %%%%%%%%%%%%%%%%%%%%%%%%%%%%%%%%%%%%%%%%%%%%%%%%%%%%%%%%%%%%%%

% \begin{thebibliography}{}

%   \bibitem[1966]{baker} Baker, N. 1966,
%       in Stellar Evolution,
%       ed.\ R. F. Stein,\& A. G. W. Cameron
%       (Plenum, New York) 333

%    \bibitem[1988]{balluch} Balluch, M. 1988,
%       A\&A, 200, 58

%    \bibitem[1980]{cox} Cox, J. P. 1980,
%       Theory of Stellar Pulsation
%       (Princeton University Press, Princeton) 165

%    \bibitem[1969]{cox69} Cox, A. N.,\& Stewart, J. N. 1969,
%       Academia Nauk, Scientific Information 15, 1

%    \bibitem[1980]{mizuno} Mizuno H. 1980,
%       Prog. Theor. Phys., 64, 544
   
%    \bibitem[1987]{tscharnuter} Tscharnuter W. M. 1987,
%       A\&A, 188, 55
  
%    \bibitem[1992]{terlevich} Terlevich, R. 1992, in ASP Conf. Ser. 31,
%       Relationships between Active Galactic Nuclei and Starburst Galaxies,
%       ed. A. V. Filippenko, 13

%    \bibitem[1980a]{yorke80a} Yorke, H. W. 1980a,
%       A\&A, 86, 286

%    \bibitem[1997]{zheng} Zheng, W., Davidsen, A. F., Tytler, D. \& Kriss, G. A.
%       1997, preprint
% \end{thebibliography}

%%%%%%%%%%%%%%%%%%%%%%%%%%%%%%%%%%%%%%%%%%%%%%%%%%%%%%%%%%%%%%%
% Appendices must be placed after   \end{thebibliography}
% They will be placed automatically on a new page.
%%%%%%%%%%%%%%%%%%%%%%%%%%%%%%%%%%%%%%%%%%%%%%%%%%%%%%%%%%%%%%%

\end{document}